\documentclass[]{spie}  
\usepackage[]{graphicx,journals}

\title{CubeSats as pathfinders for planetary detection: the FIRST-S satellite} 

\author{S. Lacour\supit{a}, V. Lapeyr\`ere\supit{a}, L. Gauchet\supit{a}, S. Arroud\supit{a}, R. Gourgues\supit{a}, G. Martin\supit{b}, S. Heidmann\supit{b}, X. Haubois\supit{a} and G. Perrin\supit{a}
\skiplinehalf
\supit{a} LESIA/Observatoire de Paris, 5 place Jules Janssen, 92195 Meudon,
France \\
\supit{b} IPAG, Univ. Grenoble Alpes, CNRS, F-38000 Grenoble, France  \\
}

\authorinfo{For further information, please send correspondence to: sylvestre.lacour@obspm.fr}

\begin{document} 
  \maketitle 
\begin{abstract}
The idea behind FIRST (Fibered Imager foR a Single
Telescope) is to use single-mode fibers to
combine multiple apertures in a pupil plane as such as to synthesize a
bigger aperture. The advantages with respect to a pure imager are i)
relaxed tolerance on the pointing and cophasing, ii) higher accuracy
in phase measurement, and iii) availability of compact, precise, and
active single-mode optics like Lithium Niobate. The latter point being
a huge asset in the context of a space mission. One of the problems of
DARWIN or SIM-like projects was the difficulty to find low cost pathfinders
missions. But the fact that Lithium Niobate optic is small and compact
makes it easy to test through small nanosats missions. Moreover, they are
commonly used in the telecom industry, and have already been tested
on communication satellites. The idea of the FIRST-S demonstrator is to
spatialize a 3U CubeSat with a Lithium Niobate nulling interferometer. The
technical challenges of the project are: star tracking, beam combination, and
nulling capabilities. The optical baseline of the interferometer would be
30\,cm, giving a 2.2\,AU spatial resolution at distance of 10\,pc. The scientific
objective of this mission would be to study the visible emission of exozodiacal light in 
the habitable zone around the closest stars. 
\end{abstract}



\section{INTRODUCTION}
\label{sec:intro}  

For direct detection and characterization of exoplanets, a space observatory is inescapable. There are two main reasons for that:
\begin{enumerate}
\item there is not enough photon on the ground within a coherent volume for proper cophasing of a telescope or interferometer. Within a coherence time of $t_0$ and within a coherence patch of diameter $r_0$, the number of photons is limited to the volume $2 \pi \,t_0\times r_0^2$. Across the whole visible band, for a bright target a magnitude 5, this volume contains nearly 5000 photons. Thus, the sole photon noise prevent any cophasing below $1/\sqrt{\rm 5000}\approx0.01\,$radian. Far from the required $10^{-5}$ radians accuracy to detect a planet with a contrast ratio of $10^{10}$.
\item only from space will we have access to optical bandpass uncontaminated by our own atmosphere. Moreover, in the mid-infrared, background thermal emission is a severe limitation to the signal to noise.
\end{enumerate}

So if space is a required passage, the remaining question is: why are we not already there? Why do we not  have a mission scheduled with the task of directly observing the photons of extrasolar planets? It is not because of the lack of --dead--  spatial projects: DARWIN, TPF-I, TPF-C, SIM, etc. One of the possible answer can be found in the 2010 decadal survey of the US\cite{Decadal}.  The authors decided to go against recommending the SIMLite mission, and gave this advice: "\textit{For the direct detection mission itself, candidate starlight suppression techniques  should be developed to a level such that mission definition for a space-based planet imaging and spectroscopy mission could start late in the decade in preparation for a mission start early in the 2020 decade.}"

However, even if the maturity of the techniques has now improved, the complexity of such a mission is still  too complex for space agencies. Nano-satellites, and more specifically CubeSats, may be the way to convince spacial agencies that the technique can be mastered in space\cite{Kjetil}. The goal of the FIRST-S project falls in that category.

\section{SCIENCE OBJECTIVE: EXOZODIACAL DUST IN THE HABITABLE ZONE}
\label{sec:base}

The FIRST-S nano-satellite is an astrophysical project, above from the fact that it is a pathfinder for spatial stellar interferometry.  The astrophysical objective is the exo-zodiacal light. The idea is to characterize the level of scattered light in the habitable zone so as to determine the amount of dust, and how it could hamper direct detection of planets. 

Easily observable in our own solar system, we also know that zodiacal dust exists in other stellar systems. Cold dust can be apparent in the form of debris disk at several tens of AU (e.g. the two most famous: Beta Pictoris and Fomalhaut). Thermal emission of warm dust have been observed with long baseline interferometers at a few tenth of astronomical units of their star\cite{2012SPIE.8445E..0XA}. But none have been observed in the habitable zone. At this distance of the star, the dust is not warm enough to be seen by its thermal signature. The stellar light scattered by the grains of dust dominate. However, the flux is expected to be several tens of thousand time fainter than the starlight.  Still, integrated over the exoplanetary system, it may be several thousand times brighter than the flux emitted by the planets.

This is why it needs a dedicated project. The problem from the ground is not the angular resolution. A 8-meter telescope theoretically reach an angular resolution equivalent to 0.12\,AU at visible wavelength (for a star at 10\,pc). The problem is the dynamic range, which is limited by the atmosphere, even with an active optics system. Our goal is therefore to reach high dynamic range from space, at moderate resolution, and in the visible (where scattering is more efficient).

The target list is based on a list of close and bright main sequence stars toward which warm dust was resolved with long baseline interferometry. It does not mean that these stars will necessarily have large amount of dust inside the habitable zone. But it gives a list of favorable targets. These stars come from a survey of 42 nearby main sequence stars made by Absil et al.\cite{2013A&A...555A.104A} at the CHARA interferometer. A third of them showed resolved emission compatible with warm dust close to the star. They serve as the basis of this work.

\begin{table}[h]
\caption{Target List} 
\label{tab:fonts}
\begin{center}       
\begin{tabular}{|l|l||l||l||l||c||c||c|} 
\hline
\rule[-1ex]{0pt}{3.5ex}  N$^0$	& Identifier	&RA	&DEC	&Sp. type&Dist. (pc) &Mag V & Photons/s\supit{a}	\\
\hline
\rule[-1ex]{0pt}{3.5ex} 1	& tau Cet	& 01 44 04.08338	& -15 56 14.9262	&G8.5V&3.7&3.50 &$1.1$	\\
\hline
\rule[-1ex]{0pt}{3.5ex} 2	& 10 Tau	& 03 36 52.38323	 &+00 24 05.9829	&	F8V&14.0& 4.30&$0.5$\\
\hline
\rule[-1ex]{0pt}{3.5ex} 3	& eta Lep	& 05 56 24.29300	 &-14 10 03.7189&	F2V&14.9 &3.72	&$0.9$\\
\hline
\rule[-1ex]{0pt}{3.5ex} 4	& lam Gem	& 07 18 05.57977&	 +16 32 25.3905&A3V	&31.0&3.58	&$1.0$\\
\hline
\rule[-1ex]{0pt}{3.5ex} 5	& bet Leo	 &11 49 03.57834	& +14 34 19.4090&	A3Va&11.0&2.13	&$3.8$\\
\hline
\rule[-1ex]{0pt}{3.5ex} 6	& ksi boo	&14 51 23.37993	& +19 06 01.6994&G7V	&6.7&4.59	&$0.4$\\
\hline
\rule[-1ex]{0pt}{3.5ex} 7	& kap CrB	 &15 51 13.93127	& +35 39 26.5671&	K0III-IV&30.6& 4.82	&$0.3$\\
\hline
\rule[-1ex]{0pt}{3.5ex} 8	& alf Lyr	 &18 36 56.33635	& +38 47 01.2802&	A0Va&7.7&0.03	&$27$\\
\hline
\rule[-1ex]{0pt}{3.5ex} 9	& 110 Her	 &18 45 39.72570	& +20 32 46.7171&	F6V&19.2&4.19	&$0.6$\\
\hline
\rule[-1ex]{0pt}{3.5ex} 10	& zet Aql	& 19 05 24.60802	& +13 51 48.5182&	A0IV-Vnn&25.5&2.99	&$2.4$\\
\hline
\rule[-1ex]{0pt}{3.5ex} 11	& alf Aql	 &19 50 46.99855	& +08 52 05.9563&	A7Vn&5.1&0.76	&$14$\\
\hline
\rule[-1ex]{0pt}{3.5ex} 12	 &alf Cep	& 21 18 34.77233	& +62 35 08.0681&A8Vn	&15.0&2.46	&$2.8$\\
\hline
\rule[-1ex]{0pt}{3.5ex} 13	 &eps Cep	 &22 15 02.19530	& +57 02 36.8771&F0IV	&26.2&4.19	 &$0.6$\\
\hline 
\end{tabular}\\
\supit{a} expected detected photon in the R band, assuming two clear circular apertures of 2\,cm, a total transmission of 10\%, and an interferometric extinction of $10^4$.
\end{center}
\end{table}

\newpage
\section{DESCRIPTION OF THE FIRST-S NANOSAT}
\label{sec:first}

   \begin{figure}
   \begin{center}
   \begin{tabular}{c}
   \includegraphics[height=7cm]{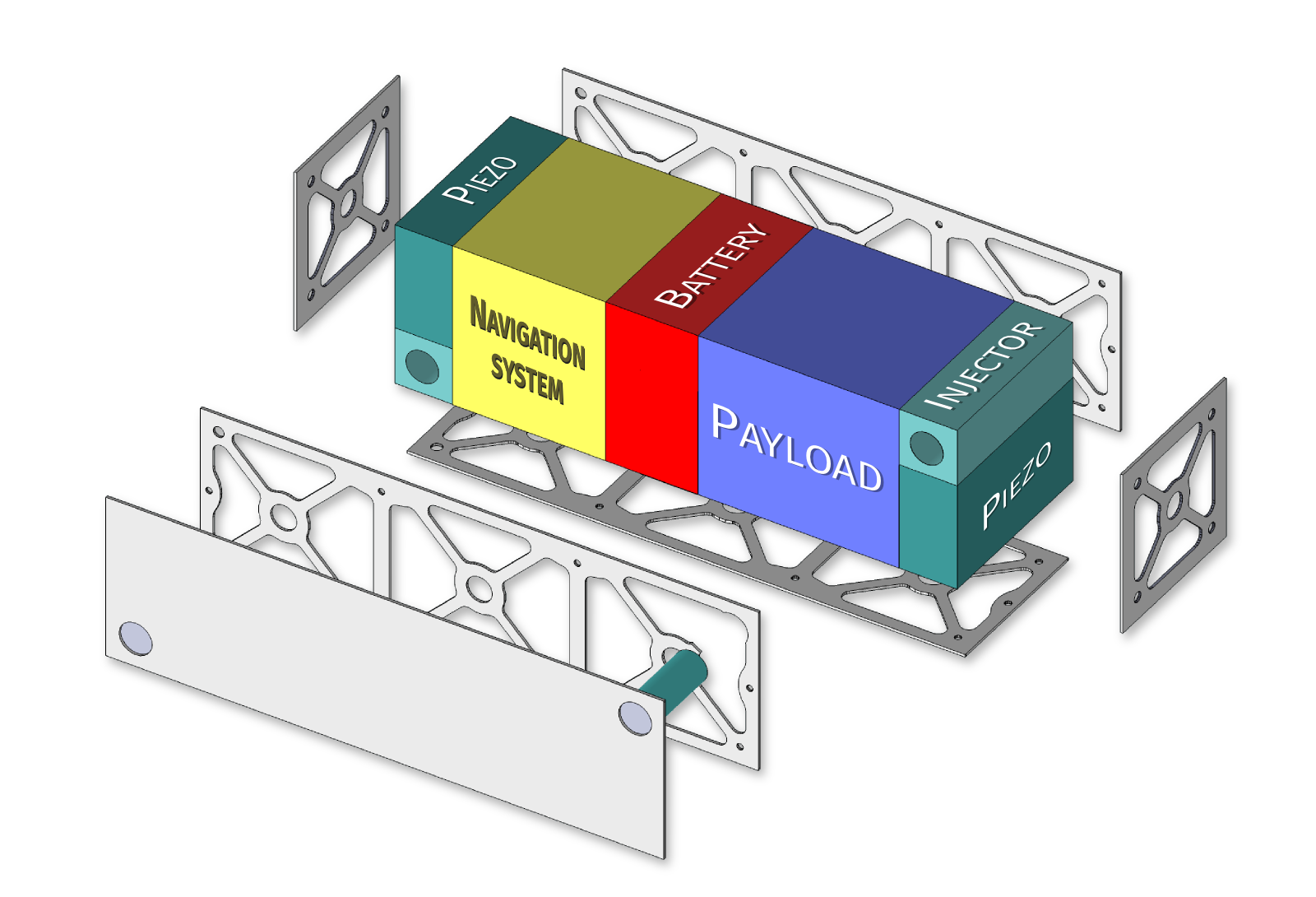}
   \end{tabular}
   \end{center}
   \caption[example] 
   { \label{fig:sat} 
Design of the FIRST-S 3U CubeSat}
   \end{figure} 

The FIRST-S CubeSat will be a 30\,cm stellar interferometer. It will have broadband nulling capabilities of the order of $10^4$ to reach a dynamic range of $10^6$. The satellite will observe the scattered dust in the R band ($\lambda = 640\,$nm). The angular resolution of the interferometer will be $\lambda/2B=220\,$mas (i.e. 2.2\,AU at 10\,pc).

The interferometer will fit in a 3U CubeSat structure. The stellar light will be received by two injectors situated on the two opposite corners of one of the lateral side of the CubeSat (the two injector modules are represented in green in figure~\ref{fig:sat}). It gives the maximum baseline length of 30\,cm without any deployable mechanism.
Each injector is made of an achromatic doublet of 2\,cm diameter feeding a single mode fiber fixed on a 3 axis piezo stage. The focal length of the lenses will be 8\,cm ($f/D=4$). Active control of the piezo will be done by the mean of a position-sensitive diode (PSD), also mounted on the piezo stage.

The core device will be an optical modulator in Lithium Niobate (LiNbO$_3$). The active integrated optics (IO) device will be in charge of i) sub-lambda phasing ii) amplitude control and iii) combination of the beams. To reach broadband capabilities of $10^4$, the IO will have to adjust the differential phase within 0.01 radian over the full wavelength bandpass. It will also have to adjust the amplitude ratio between the two beams within 1\% (also over the full bandpass). Finally, it will combine the two beams with a Y-junction. The exit output of the IO will be a simple single-mode fiber.

The detection will be done thanks to a single photon counting avalanche photodiode (SAPD). The advantage of photon counting is the possibility to do optical modulation at any frequency, followed by demodulation, without any readout noise. The only limitation is the dark-current rate (DCR), typically of the order of 10 photons per second in the visible. This 10 photons per second will be the background limitation for the nulling. Assuming that we can reach an extinction of $10^4$, it means that the signal to noise of all the stars with less than $10^5$ photons per second will be limited by the DCR. We are currently looking for SAPD with lower DCR.

In the last column of Table~1 is an estimation of the stellar flux during normal operation (to be compared with the DCR). The values correspond to the flux of photon that we expect in the R band, detected through two 2\,cm apertures during one second. It includes a factor $10^4$ due to the interferometric nulling, plus a decrease by a factor 10 caused by the CubeSat transmission. This 10\% includes i) 70\% to inject the light into the single mode fiber, ii) 50\% due the use of only one polarisation, iii) 50\% which is the expected transmission of the LiNbO$_3$ device (including coupling into it), and iv) 70\% which correspond to the quantum efficiency of the SAPD. 

\textit{Summary} : - Interferometric nulling: extinction 10$^4$ (beams must be phased at 1\,nm level)
- Dynamic range 10$^6$ (in one hour on Alpha Cep)
- Resolution: 220\,mas (2.2 \,AU at 10\,pc)
- Wavelength: R band ($\lambda_o = 640\,$nm, $\Delta \lambda/\lambda= 0.23$)
- Polarisation: just one, the linear polarization along the baseline axis.

\section{NULLING WITH LITHIUN NIOBATE ACTIVE OPTICS }
\label{sec:niobate}

   \begin{figure}
   \begin{center}
   \begin{tabular}{c}
   \includegraphics[height=6cm]{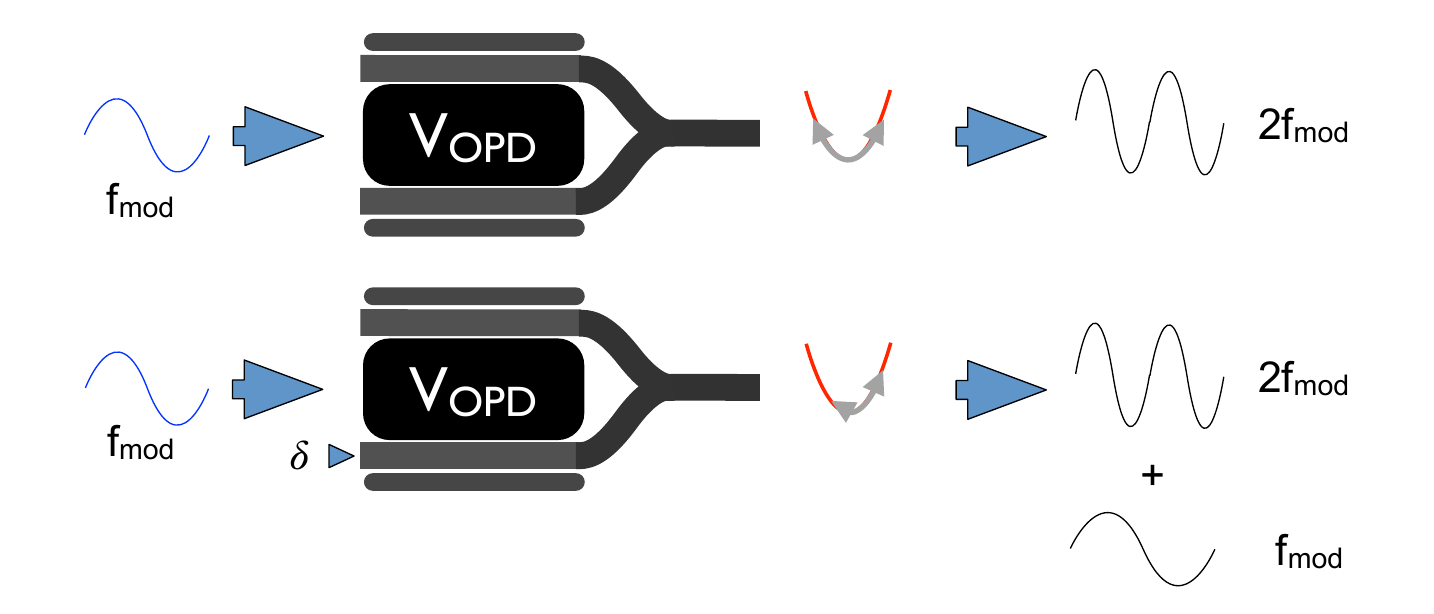}\\
   \includegraphics[height=2.5cm]{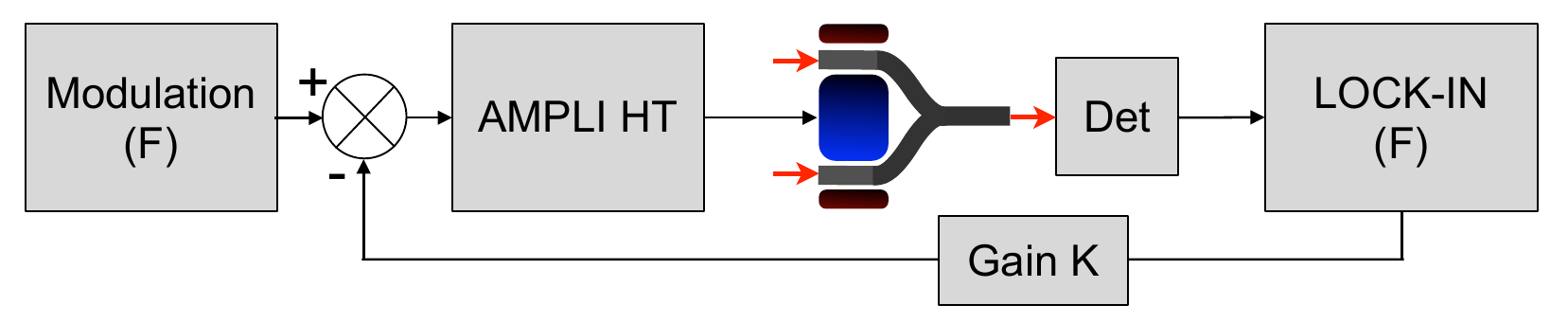}
   \end{tabular}
   \end{center}
   \caption[example] 
   { \label{fig:heidmann} 
Close loop system to track the null. The OPD is modulated thanks to a voltage applied on the electrodes. The flux at the output of the IO is then modulated at the same frequency if it is away from the null. This information is then used in closed loop (lower panel) to adjust the mean voltage.\cite{theseH}}
   \end{figure} 

LiNbO$_3$ is an asymmetric crystal with many properties. It is piezoelectrique, pyroelectrique, and ferroelectrique.
 But for our use, the most fundamental characteristic is that the material is electro-optic: the refractive index of the material changes when subjected to an electric field. This electric field may come from an electric voltage applied between two electrodes on each side of a waveguide. The first order effect is a linear electro-optic effect, or the Pockels effect, where the refractive index of the material varies in proportion to the electric field applied. The importance of the Pockels effect is that it can be used to control the phase and amplitude (thanks to a Mach-Zehnder) of a signal inside a waveguide. 
 
 We will use that effect to control the phase and amplitude of the light coming from the two apertures. Figure~\ref{fig:heidmann} presents the method that we will use for nulling. The voltages on the electrodes will be modulated at a frequency of the order of the kHz: for example, 2kHz for the differential phase, 1kHz for the differential amplitude, and between 500 and 900 Hz for each one of the electrodes controlling the dispersion. The SAPD will detect the signal at high frequency (thanks to a gate time of the order of 1\,ns, or 1\,MHz in frequency) and will do synchronous deconvolution of the signal for each controlled parameter. The deconvolved signal is then added as a constant to each of the modulated signal sent to the electrodes.

\section{ATTITUDE \& OPD, SENSING \& CONTROL}
\label{sec:attitude}

The specificity of the mission -- interferometric nulling with single-mode fibers -- requires high precision control of the spacecraft. These contraints cannot be considered independently for the scientifique payload and for the platform. They satellite has to be considered as a whole. The requirements are of two types:
\begin{enumerate}
\item The stellar light must be injected into the fiber. The core of the single-mode fiber is of the order of 3$\,\mu$m in diameter. As a result, considering the focal length of 8\,cm, it means that the field must be controlled well below 1 arcminute to have a proper injection. 
\item The OPD between the two arms of the interferometer must be controlled up to 1\,nm. The LiNbO$_3$  active IO will ensure fine tuning of the beams. But the optical delay that the IO can induce is short, only of the order of a few microns for 10 volts applied on the electrode. \end{enumerate}

To fulfill these requirements, the spacecraft will use three different layers for attitude and OPD determination:  i) an Earth Horizon Sensor (EHS) camera , for coarse attitude determination, ii) positioning-sensitive diodes (PSD), and iii) interferometric sensing thanks to phase modulation of the beams (see figure~\ref{fig:heidmann}).

The spacecraft will also need three different layers of attitude and OPD control: i) the classical reaction wheels and magnetic bars, ii) XYZ piezo stages on which are mounted the fibers, and iii) the LiNbO$_3$ active IO.

The overall combination of devices, to get to a final OPD budget of 1\,nm, and attitude positioning of 1 arc second, is summarized in the following table:

\begin{table}[h]
\caption{OPD and Attitude: the different layers of detection and correction} 
\label{tab:fonts}
\begin{center}       
\begin{tabular}{|l|c|c|c|c|} 
\hline
\multicolumn{5}{|c|}{\bf Sensing} \\
\hline
\rule[-1ex]{0pt}{3.5ex}     & \multicolumn{2}{c|}{Field ($\alpha$,$\delta$)} & \multicolumn{2}{c|}{OPD} \\
\rule[-1ex]{0pt}{3.5ex}     & range & accuracy & range & accuracy \\
\hline
\rule[-1ex]{0pt}{3.5ex}   Earth Horizon Sensor Camera & $\pm 180^o$ & $\pm 0.2^o$ & $inf$  & $\pm 0.5\,$mm\supit{a} \\
\hline
\rule[-1ex]{0pt}{3.5ex}   Position-Sensitive Diode& $\pm 1.4^o$ \supit{b}& $\pm 1.3$'' \supit{b}&$\pm 4\,$mm\supit{a}   & $\pm 1\,\mu$m\supit{a}  \\
\hline
\rule[-1ex]{0pt}{3.5ex}   Interferometer & \multicolumn{2}{c|}{N/A}  & $\pm 3\,\mu$m\supit{c} & $\pm 1\,$nm \\
\hline
\hline
\multicolumn{5}{|c|}{\bf Control} \\
\hline
\rule[-1ex]{0pt}{3.5ex}     & \multicolumn{2}{c|}{Field ($\alpha$,$\delta$)} & \multicolumn{2}{c|}{OPD ($\mu$m)} \\
\rule[-1ex]{0pt}{3.5ex}     & range & accuracy & range & accuracy \\
\hline
\rule[-1ex]{0pt}{3.5ex}   Reaction wheels & $\pm 180^o$ & $\pm 1$' \supit{d} & $inf$ & $\pm 40\,\mu$m\supit{a} \\
\hline
\rule[-1ex]{0pt}{3.5ex}   XYZ Piezos & $\pm 25$' \supit{e} & $\pm 0.0025$''  & $\pm 60\,\mu$m \supit{e}& $\pm 10\,$nm \\
\hline
\rule[-1ex]{0pt}{3.5ex}   LiNbO$_3$ Active Optic & \multicolumn{2}{c|}{N/A} & $\pm 3\,\mu$m & $\pm 1\,$nm  \\
\hline 
\end{tabular}\\
\end{center}
\supit{a} Values deduced from field values assuming a 15\,cm baseline between spacecraft barycenter and optics \\
\supit{b} Physical size of PSD is 4\,mm with a positioning accuracy of $0.5\,\mu$m \\
\supit{c} The coherence length of the interferometer is limited by the bandwidth and the wavelength ($\lambda^2/\Delta \lambda$)\\
\supit{d} Sub-arcminute accuracy of CubeSat reaction wheels have already be demonstrated (Pong et al. 2011)\cite{Pong}\\
\supit{e} Piezo range is 120$\mu$m
\end{table}

\section{CONCLUSION}
\label{sec:conc}

A single-mode stellar interferometer in space is at reach, even though it may be risky because of the complexity of the system. However, in this context of high risk/high gain mission, CubeSats offer an ideal environment. Simply put: risks can be taken thanks to the low cost of the mission.

Nano-satellites are more than just hype. They are more than cool tools to motivate students. They offer formidable opportunities for astronomical missions. But the way space missions are designed has to be rethought and adapted. Traditional space mission preparations cannot apply. Also, CubeSats cannot do the work of larger space missions. But by rethinking the projects in the context of nano-satellites, formidable new ideas of missions can emerge. As well as groundbreaking astrophysical results.


 


\bibliography{lithium}   
\bibliographystyle{spiebib}   

\end{document}